\documentclass[12pt]{article}

\usepackage{epsf,graphicx}
\usepackage{latexsym}

\makeatletter
\newcommand\eqsecnum{
\@newctr{equation}[section]
\renewcommand\theequation{\arabic{section}.\arabic{equation}}%
}
\newcommand{\vereq}[2]{\lower3pt\vbox{\baselineskip1.5pt \lineskip1.5pt
\ialign{$\m@th#1\hfill##\hfil$\crcr#2\crcr\sim\crcr}}}

\renewcommand\appendix{\par
  \setcounter{section}{0}%
  \setcounter{subsection}{0}%
  \gdef\thesection{\@Alph\c@section}%
\renewcommand\theequation{\Alph{section}.\arabic{equation}}
}
\newcount\@indentflag \global\@indentflag=1 %
\newdimen\@eqtoeqnum \@eqtoeqnum=6pt %
\def\@indentamount{%
\ifcase\@indentflag 0pt\or\@centering\or0pt plus1fil\fi\relax
}
\def\FL{\global\@indentflag=0 }
\def\FR{\global\@indentflag=2 }
\def\@eqnnum{\hbox{\reset@font\rm(\theequation)}}
\let\make@eqnnum=\@eqnnum %
\def\eqnum#1{\dec@eqnnum \global\def\make@eqnnum{\reset@font\rm(#1)}%
\def\@currentlabel{#1}%
}
\def\inc@eqnnum{\addtocounter{equation}{1}}
\def\dec@eqnnum{\addtocounter{equation}{-1}}
\newbox\@testboxa
\newbox\@testboxb
\def\equation{\par\vskip-\lastskip\vskip\abovedisplayskip
\inc@eqnnum\let\@currentlabel=\theequation
\setbox\@testboxa=\hbox\bgroup\hskip\@totalleftmargin\hskip\@indentamount
\hbox\bgroup$\displaystyle
}
\def\endequation{$\egroup\hskip\@centering\egroup %
\setbox\@testboxb=\hbox{\make@eqnnum}%
\bgroup
\@tempdima\wd\@testboxa \advance\@tempdima by\wd\@testboxb
\ifcase\@indentflag
\advance\@tempdima by\@eqtoeqnum
\ifdim\@tempdima<\hsize %
\def\@tempa{0}%
\else
\def\@tempa{1}%
\fi
\or
\advance\@tempdima by2\@eqtoeqnum
\ifdim\@tempdima<\hsize %
\def\@tempa{0}%
\else %
\@tempdima\wd\@testboxa \advance\@tempdima by\wd\@testboxb
\advance\@tempdima by\@eqtoeqnum
\ifdim\@tempdima<\hsize %
\def\@tempa{0}%
\setbox\@testboxa\hbox{\hfill\box\@testboxa\kern\@eqtoeqnum}%
\else
\def\@tempa{1}%
\fi
\fi
\or
\advance\@tempdima by2\@eqtoeqnum
\ifdim\@tempdima<\hsize %
\def\@tempa{0}%
\setbox\@testboxb=\hbox{\kern\@eqtoeqnum\make@eqnnum}%
\else
\def\@tempa{1}%
\fi
\fi
\ifnum\@tempa=0 %
\hbox to\hsize{\unhbox\@testboxa\box\@testboxb}%
\else %
\vbox{\hbox to\hsize{\unhbox\@testboxa}%
\vskip6pt %
\hbox to\hsize{\hfil\box\@testboxb}}%
\fi
\egroup
\global\let\make@eqnnum\@eqnnum %
\vskip\belowdisplayskip\noindent\global\@indentflag=1 \global\@ignoretrue
}
\def\eqnarray{\par\vskip-\lastskip\vskip\abovedisplayskip
\inc@eqnnum\let\@currentlabel=\theequation
\global\@eqnswtrue\m@th
\global\@eqcnt\z@
\tabskip\@totalleftmargin\advance\tabskip by\@indentamount\let\\\@eqncr
\halign to\hsize\bgroup\hskip\@centering
$\displaystyle\tabskip\z@{##{}}$&\global\@eqcnt\@ne
\hfil${{}##{}}$\hfil
&\global\@eqcnt\tw@ $\displaystyle\tabskip\z@{##}$\hfil
\tabskip\@centering \if@eqnsw\phantom{\make@eqnnum\kern\@eqtoeqnum}\fi
&\llap{##}\tabskip\z@\cr}
\def\endeqnarray{%
\@@eqncr\egroup
\vskip\belowdisplayskip\noindent
\dec@eqnnum\global\@indentflag=1
\global\let\make@eqnnum\@eqnnum %
\global\@ignoretrue
}
\newbox\tempboxa
\newdimen\captionboxsubcount 
\def\capsize#1{\captionboxsubcount=#1pt}
\newdimen\captionboxsub
\captionboxsub=\hsize \advance\captionboxsub by -\captionboxsubcount
\advance\captionboxsub by -\captionboxsubcount
\long\def\@makecaption#1#2{
 \setbox\@tempboxa\hbox{\footnotesize #1: #2}
 \ifdim \wd\@tempboxa >\captionboxsub 
\rightskip=\captionboxsubcount \leftskip=\captionboxsubcount 
  \footnotesize #1: #2 
\else \hbox to\hsize{\hfil\box\@tempboxa\hfil} 
 \fi}

\makeatother

\capsize{30}

\eqsecnum

\renewcommand{\today}{\ifcase\month\or
 Jan.\or Feb.\or Mar.\or Apr.\or May\or Jun.\or
 Jul.\or Aug.\or Sep.\or Oct.\or Nov.\or Dec.\fi
 \space\number\day, \number\year}

\begin{document}

\thispagestyle{empty}

\vspace*{0.5cm}

\begin{flushright}
\begin{minipage}{4cm}
\begin{flushleft}
DPNU-05-19
\end{flushleft}
\begin{flushleft}
SU-4252-820
\end{flushleft}
\end{minipage}
\end{flushright}

\begin{center}
{\bf\Large
Chiral approach to Phi Radiative Decays 
}
\ \\ 
\end{center}

\vspace{0.5cm}

\begin{center}
\large
Deirdre {\sc Black}$^{\rm(a)}$,
Masayasu {\sc Harada}$^{\rm(b)}$ and
Joseph {\sc Schechter}$^{\rm(c)}$
\end{center}

\begin{flushright}
\begin{minipage}{14cm}
\begin{flushleft}
$^{\rm(a)}$ {\it University of Cambridge, Department of Physics
Cavendish Laboratory,}\\
\qquad {\it J J Thomson Avenue
Cambridge CB3 0HE, UK}
\\
$^{\rm(b)}$ {\it Department of Physics,
Nagoya University, Nagoya 464-8602, Japan} \\
$^{\rm(c)}$ {\it Department of Physics, Syracuse University
Syracuse, NY 13244-1130, USA}
\end{flushleft}
\end{minipage}
\end{flushright}

\vspace{0.5cm}

\begin{abstract}
The radiative decays of the phi meson are known to be a
 good source of information about the $a_0$(980) and $f_0$(980)
scalar mesons. We discuss these decays starting from a 
non-linear model Lagrangian which maintains the (broken) chiral
symmetry  
for the pseudoscalar (P), scalar (S) and vector (V) nonets involved.
The characteristic feature is derivative coupling for the SPP
interaction. In an initial approximation which models
 all the scalar nonet 
radiative processes together with the help of a point like vertex,
it is noted that the derivative coupling prevents the $a_0$
and $f_0$ resonance peaks from getting washed out (by falling
phase space). However, the shapes of the invariant two final PP
 mass distributions do not agree well with the experimental
ones. For improving the situation
 we verify that inclusion of the charged K 
meson loop diagrams
in the model does reproduce the experimental spectrum
shapes in the resonance 
region. The derivative coupling introduces quadratic as well as 
logarithmic divergences in this calculation. Using dimensional 
regularization we show in 
detail that these divergences actually cancel out
among the four diagrams, as expected
from gauge invariance. We point out the features which are
expected to be important for further work on this model and for
learning  
more about the puzzling scalar mesons. 
\end{abstract}

\newpage

\section{Introduction}

    Recently, there have been a number of important experimental
 studies \cite{recentexpts} 
of the rare radiative decays of the $\phi$(1020) vector meson:
$\phi \rightarrow \pi \pi \gamma$ and $\phi \rightarrow \pi
\eta\gamma$. These decays seem to be dominated by the
production (and subsequent decay) of the scalar mesons,
$a_0$(980) and $f_0$(980) according to $\phi \rightarrow
f_0, a_0 +\gamma$ and hence are generally considered to
provide valuable information about the puzzling light scalar
mesons\cite{scalars} of low energy QCD.

    The theoretical analysis of this type of decay was
initiated by Achasov and Ivanchenko \cite{Achasov-Ivanchenko} and
followed 
up by many others \cite{radiative}. The starting point was the
observation 
that the $\phi$ meson decays about 50 per cent of the time
into ${K^+K^-}$. Since this final state can easily annihilate
to produce either an $f_0$ or $a_0$ together with an emitted photon,
it is rather natural to consider charged $K$-meson loop diagrams
to describe the process. Similarly the $\phi$ meson is observed to
decay 
about 15 per cent of the time to $\pi\rho$ or $\pi^+\pi^-\pi^0$
so one expects some non resonant background which is likely to include
the emission of a pion with a virtual $\rho$ which subsequently
decays into  $\pi\gamma$ (and similar diagrams leading to a 
$\pi^0\eta \gamma$ final state).

     The varied calculations along these lines lead to
results which more or less agree with experiment. Of course it
is desirable to fine tune this agreement, both to reflect
the expected improved accuracy of new experiments as well
as to improve our understanding of strong interaction calculations.
Here we will focus on some technical points, which do not
much change the previous results but may be of interest for future
more ambitious calculations as more experimental
data become available.
 Mainly, we will require that the amplitudes
all be computed from a chiral invariant Lagrangian (containing usual
quark mass induced breaking terms). This is a symmetry of nature 
apparently so it is desirable to calculate in this way even though the
spontaneous breakdown of chiral symmetry (in the absence of quark mass
terms) means that, especially away from thresholds, one can often get
reasonable predictions by not explicitly taking it into account.

    Two approaches are commonly employed to implement the chiral
symmetry in the effective Lagrangian framework. In the linear
sigma model approach, scalar partners of the pseudoscalars
are introduced. In the non-linear sigma model approach, one
initially deals with pseudoscalars only, the scalars having been
essentially 
``integrated out". The characteristic feature of the non linear model
is the appearance of derivative type interaction terms as opposed
to non derivative type interaction terms in the linear
model. Nevertheless, 
the non linear model is often more convenient to use. For example,
the celebrated result \cite{weinberg} for near threshold pi pi
scattering  arises in the 
linear model from a delicate cancellation of two rather large
terms. On the other hand it arises directly from a simple single
term of the correct characteristic strength in the non linear model.
In the present paper we shall deal with the non linear model
approach. Since vector and scalar mesons are also involved in the
processes 
of interest we will add these to the non linear Lagrangian of
pseudoscalars 
in a conventional way. Such a formulation essentially implements
vector meson dominance automatically for processes involving photons. 

   We shall restrict our attention further here to processes of
the type $\phi \rightarrow \gamma$ + virtual scalar where the virtual
scalar 
(either $a_0$ or $f_0$) subsequently decays to two
pseudoscalars. First we shall consider a possible non - $K^+$ loop
contribution to this process. We previously \cite{BHS} studied this by
introducing an effective strong 
 VVS (vector-vector scalar) interaction based on an analogy to the
 effective VVP (vector vector
pseudoscalar) interaction used many years ago~\cite{gsw} to 
study analogous processes
like $\omega$(782) $\rightarrow \pi^0 \gamma$. This
might open the possibility
of understanding properties of the whole nonet of scalars
at once. 
Especially, it might shed some light on the composition of the
light scalar nonet; whether the light scalar mesons are
composed of one quark and one anti-quark (2-quark picture)
or two quarks and two anti-quarks (4-quark picture).

 In the present paper,
 we point out an interesting effect.
If a non derivative SPP type interaction were to be used there would
be a strong tendency for the decreasing phase space to wash out 
 the predicted scalar meson peak in, for example, 
$d \Gamma(\phi\rightarrow \pi^0\eta\gamma)/dq$. Here
$q^2$ is the invariant squared mass for the $\pi \eta$ 
system. On the other hand, the use of a derivative type SPP
interaction, as is required for chiral symmetry in the
non linear sigma model approach, restores the peak.
 (There is not 
necessarily any contradiction with the expectation that the same
 physics 
near threshold should be expressed by suitably generalized 
linear and non linear models. One expects the linear model description to 
include additional terms).
However, we notice that there is experimentally more enhancement
 of the scalar peak than can be accounted for by this mechanism.
Thus we are led to also consider the  usual $K^+$ loop diagram  
in our approach.  

    As mentioned, the $K^+$ loop diagram has been considered
by many authors~\cite{Achasov-Ivanchenko,radiative}. We can not
basically  
change the well
established results. However we note that the effect of
the derivative couplings will also sharpen the scalar peak.
Actually, the derivative SPP couplings result in  quadratic
as well as logarithmic divergences
and an additional diagram. It has been found~\cite{quadratic}
 that such``unpleasant details" 
of the calculation can be circumvented by assuming gauge 
invariance. Specifically, gauge invariance requires that the
amplitude for $\phi \rightarrow$ photon + scalar be proportional to
$\epsilon_\mu \epsilon^V_\nu(\delta_{\mu\nu}p{\cdot}k
-p_\mu k_\nu)$, where $\epsilon^V$ and $p$ are respectively the
polarization 
and momentum four vectors of the $\phi$ meson while $\epsilon$
and $k$ correspond to the photon. Then it is only necessary to
calculate 
the coefficient of the $p_\mu k_\nu$ term, which eliminates
the need to calculate two diagrams and worry about divergences
actually cancelling each other. Of course it would be nice to
regulate all the diagrams and verify in detail
 how the cancellations take place. We have carried out this somewhat 
lengthy task using the dimensional regularization scheme
and will give details in the present paper.

%   Finally, we will give some fits to the experimental data based
%on including both chiral invariant diagrams of the K-loop type
%and VVS contact interaction type. These fits are actually reasonably
%good 
%but must be considered as illustrative since no explicit
%background (ie, non-resonant) contributions were added.

In section~\ref{sec:VVS}, 
we first present the chiral Lagrangian of
pseudoscalars, vectors and scalars which will be used for the
subsequent calculations. Our initial motivation, described 
in Ref.~\cite{BHS}, was to relate all the decays of the types 
S$\rightarrow\gamma\gamma$, 
V $\rightarrow$S$\gamma$ and  S $\rightarrow$V$\gamma$
to each other by using a simple effective point like interaction.
We next consider the $\phi(1020)$ decays into $\pi^0\eta$
and $\pi^0\pi^0$ proceeding respectively from intermediate
$a_0(980)$ and $f_0(980)$ resonances in this simple model. It 
can be seen that the spectrum shapes for large $q$ are not as
sharply peaked as the experimental data indicate.

   In section~\ref{sec:K-loop}, 
we calculate the form of the charged $K$ meson loop 
contributions to these two decays using a non-linear chiral Lagrangian 
which maintains the chiral invariance when vectors and scalars as
well as pseudoscalars are included. The extension to include 
photon interactions is given. It is noted that individual diagrams
contain quadratic as well as logarithmic divergences. A careful
treatment using the dimensional regularization scheme shows that
these divergences both cancel leaving a finite answer. 
  
   In section~\ref{sec:comp} 
we study the spectrum shape of the $K$-loop
contributions to these decays. We find that this has a
 characteristic shape which does in fact agree with experiment,
 suggesting that the dynamics of the $K$ loop plays
an important role.

    Section~\ref{sec:SD} contains a brief summary. Some discussion
will be given on the status of the present program and related
future work.

\section{VVS type contributions to $\phi\rightarrow \pi^0\eta\gamma$
and $\phi\rightarrow \pi^0\pi^0\gamma$}
\label{sec:VVS}

    Our calculation is based on a standard non-linear chiral 
Lagrangian
containing, in addition to the pseudoscalar nonet matrix field $\phi$,
the vector meson nonet matrix $\rho_\mu$ and a scalar nonet matrix
field denoted by $N$.
Under chiral unitary transformations of the three light quarks;
$q_{\rm L,R} \rightarrow U_{\rm L,R} \cdot q_{\rm L,R}$, the chiral
matrix $U = \exp ( 2 i \phi/F_\pi)$,
where $F_\pi \simeq 0.131\,\mbox{GeV}$, transforms as 
$U \rightarrow U_{\rm L}\cdot U \cdot U_{\rm R}^\dagger$.
The convenient matrix
$K(U_{\rm L}, U_{\rm R}, \phi )$~\cite{CCWZ}
is defined by the following transformation property of 
$\xi$ ($U = \xi^2$):
$\xi \rightarrow U_{\rm L} \cdot \xi \cdot K^{\dag} 
  = K \cdot \xi \cdot U_{\rm R}^{\dag}$,
and specifies the transformations of ``constituent-type'' objects.
The fields we need transform as
\begin{eqnarray}
&&
  N \rightarrow K \cdot N \cdot K^{\dag} \ ,
\nonumber\\
&&
  \rho_\mu \rightarrow K \cdot \rho_\mu \cdot K^{\dag}
  + \frac{i}{\tilde{g}} K \cdot \partial_\mu K^{\dag}
\ ,
\nonumber\\
&&
  F_{\mu\nu}(\rho) = 
  \partial_\mu \rho_\nu - \partial_\nu \rho_\mu - i 
  \tilde{g} \left[ \rho_\mu \,,\, \rho_\nu \right]
  \rightarrow
  K \cdot F_{\mu\nu} \cdot K^{\dag}
\ ,
\label{transf}
\end{eqnarray}
where the coupling constant $\tilde{g}$ is about $4.04$.
One may refer to Ref.~\cite{Harada-Schechter} for our treatment of the
pseudoscalar-vector Lagrangian and to
Ref.~\cite{Black-Fariborz-Sannino-Schechter:99} for the scalar
addition.
The entire Lagrangian is chiral invariant (modulo the quark mass term
induced symmetry breaking pieces) and, when
electromagnetism is added, gauge invariant. The $U(3)_L\times U(3)_R$
invariant portion of the effective Lagrangian reads:~\footnote{%
 This Lagrangian can be rewritten within the framework of 
 the hidden local symmetry (HLS)~\cite{BKY:88,HY:PRep}.
 The method of including vector mesons used in this paper
 based on the proposal in Refs.~\cite{KRS:KS} is equivalent
 to that based on the HLS approach at tree level.~\cite{equivalence}
 When we consider the vector mesons inside the loop,
 the two approaches might have some differences.
 In the present analysis, however, we will consider the
 loop corrections from only the kaon, which provides a large
 enhancement to the $\phi$ radiative decay amplitude.
 All other loop corrections from vector mesons are naturally 
 expected to be small.
 In this sense, the method used in this paper is completely equivalent
 to the recently developed method~\cite{HY:PRep} used in the HLS.
 Note that the scalar mesons have not been included inside the loop in 
 either approach. 
}
\begin{eqnarray}
{\cal L}_0 = &-& \frac{F_\pi^2}{2} {\rm Tr} (p_\mu p_\mu)
 - \frac{1}{4} {\rm Tr} ( F_{\mu \nu}(\rho) F_{\mu
\nu}(\rho) ) \nonumber \\
& - & \frac{1}{2} {\rm Tr} ( {\cal D}_\mu N {\cal D}_\mu N ) -
\frac{m_v^2}{2 {\tilde g}^2} {\rm Tr} \left[ { ( \tilde g \rho_\mu -
v_\mu)}^2 \right] \nonumber \\
&-& a {\rm Tr} (NN) - c {\rm Tr} (N) {\rm Tr}(N) \nonumber \\
&+& F_\pi^2  [ A \epsilon^{abc} \epsilon_{def} N_a^d
{(p_\mu)}_b^e {(p_\mu)}_c^f + B {\rm Tr} (N) {\rm Tr}{(p_\mu p_\mu)}
\nonumber \\
 &+& C{\rm Tr}(Np_\mu){\rm Tr}(p_\mu) + D {\rm Tr} (N) {\rm Tr} (p_\mu)
{\rm Tr} (p_\mu)]
\,
\label{LagL0}
\end{eqnarray}
where 
%${\cal D}_\mu N = \partial_\mu N - i v_\mu N + i Nv_\mu$.
${\cal D}_\mu N = \partial_\mu N - i \tilde{g} \rho_\mu N 
   + i N \tilde{g} \rho_\mu$.\footnote{%
 One could also use for the covariant derivative, the combination
 $c{\tilde g}\rho_\mu+(1-c)v_\mu$ with $c$ being an arbitrary 
 constant.
 In any case, there are a few more terms such as
 $\mbox{tr} \left( \left(\tilde{g}\rho_\mu - v_\mu \right) N
   \left(\tilde{g}\rho_\mu - v_\mu \right) N
  \right)$ and
 $\mbox{tr} \left( \left(\tilde{g}\rho_\mu - v_\mu \right)^2 N^2
  \right)$, which include the same number of derivatives.
 We note that
 the above extra terms as well as the interaction terms from the
 covariant derivative do not contribute in the present analysis,
 where we are considering the processes related to only one
 scalar meson.
}
Furthermore $v_\mu, p_\mu =(i/2)(\xi\partial_\mu \xi^{\dagger}\pm
\xi^{\dagger}\partial_\mu \xi)$, where $\xi=U^{1/2}$. 
  These terms
include the parameters $m_v^2, a, c, A, B, C$ and $D$. More details
about 
the evaluation of these parameters are discussed 
in Refs.~\cite{abfs} and \cite{Harada-Schechter}.

It should be remarked that the effect of adding vectors to the chiral
Lagrangian of pseudoscalars only is to replace the photon coupling to
the charged pseudoscalars as,
\begin{eqnarray}
&&
i e {\cal A}_\mu \, 
\mbox{Tr}\left(
  Q \phi \mathop{\partial_\mu}^{\leftrightarrow} \phi 
\right)
\rightarrow
\nonumber\\
&& \quad
e {\cal A}_\mu \, 
\biggl[
  k \tilde{g} F_\pi^2 
  \mbox{Tr} \left( Q \rho_\mu \right)
\nonumber\\
&& \qquad
  + i \left( 1- \frac{k}{2} \right) 
  \mbox{Tr}\left(
    Q \phi \mathop{\partial_\mu}^{\leftrightarrow} \phi 
  \right)
\biggr]
+ \cdots
\ ,
\label{VMD}
\end{eqnarray}
where ${\cal A}_\mu$ is the photon field, 
$Q = \mbox{diag}(2/3,-1/3,-1/3)$ 
and $k = \left( \frac{m_v}{\tilde{g} F_\pi} \right)^2$
with $m_v \simeq 0.76\,\mbox{GeV}$.
The ellipses stand for symmetry breaking corrections.
We see that in this model, Sakurai's vector meson 
dominance~\cite{Sakurai} simply amounts to the 
statement that $k=2$ (the KSRF 
relation~\cite{KSRF}).
This is a reasonable numerical approximation which is essentially
stable to the addition of symmetry 
 breakers~\cite{Harada-Schechter,foot:VD} 
and we employ it here by neglecting the last term
in Eq.~(\ref{VMD}).

The proposed effective SVV type terms in the effective 
Lagrangian are~\cite{BHS}:
\begin{eqnarray}
&&
{\cal L}_{SVV} =  \beta_A \,
\epsilon_{abc} \epsilon^{a'b'c'}
\left[ F_{\mu\nu}(\rho) \right]_{a'}^a
\left[ F_{\mu\nu}(\rho) \right]_{b'}^b N_{c'}^c
\nonumber\\
&& \quad
{}+
 \beta_B \, \mbox{Tr} \left[ N \right]
\mbox{Tr} \left[ F_{\mu\nu}(\rho) F_{\mu\nu}(\rho) \right]
\nonumber\\
&& \quad
{}+
 \beta_C \, \mbox{Tr} \left[ N F_{\mu\nu}(\rho) \right]
\mbox{Tr} \left[ F_{\mu\nu}(\rho) \right]
\nonumber\\
&& \quad
{}+
 \beta_D \, \mbox{Tr} \left[ N \right]
\mbox{Tr} \left[ F_{\mu\nu}(\rho) \right]
\mbox{Tr} \left[ F_{\mu\nu}(\rho) \right]
\ .
\label{SVV}
\end{eqnarray}
Chiral invariance is evident from Eq.~(\ref{transf}) and the four
flavor-invariants are needed for generality.  (A term 
$\sim \mbox{Tr}( FFN )$ is linearly dependent on the four shown).
Actually the $\beta_D$ term does not contribute in our model so there
are only three relevant parameters $\beta_A$, $\beta_B$ and $\beta_C$.

\subsection{$a_0(980)$ production}
\label{ssec:1 a0}

\begin{figure}[htbp]
\begin{center}
\epsfxsize=13cm
\epsfbox{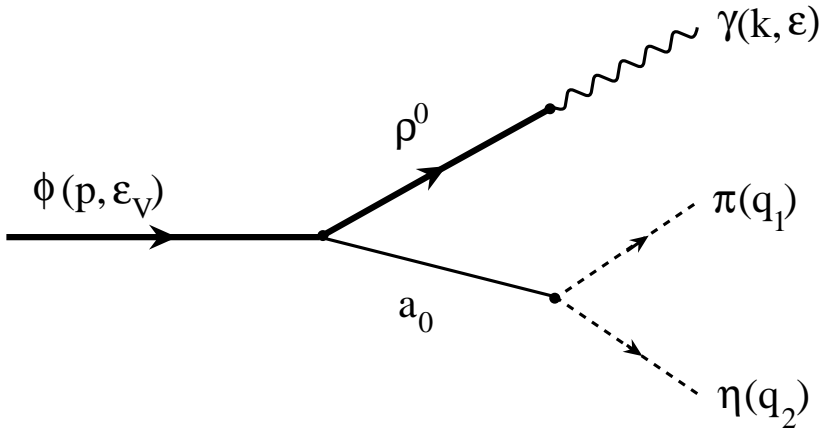}
\end{center}
\caption[]{Feynman diagram for 
$\phi(p,\epsilon^V) \rightarrow 
  \pi^0(q_1) \eta(q_2) \gamma(k,\epsilon)$ 
using an effective VVS term
}\label{fig:a0tree}
\end{figure}

  The Feynman diagram for the  contribution from the new VVS terms to
the decay process
$\phi(p,\epsilon^V) \rightarrow 
  \pi^0(q_1) \eta(q_2) \gamma(k,\epsilon)$
is shown in Fig. \ref{fig:a0tree}.
 Note that the
photon is produced through its mixing with vector mesons according to 
Eq.~(\ref{VMD}). The Feynman amplitude is

\begin{equation}
e(q_1\cdot q_2)Y^{(\pi\eta)}_{a_0}
\left[(p\cdot k)(\epsilon^V\cdot\epsilon)-
(p\cdot\epsilon)(k\cdot\epsilon^V)
\right]
\ ,
\end{equation}
where
\begin{eqnarray}
&&
Y_{a_0}^{(\pi\eta)} =
\frac{C_\phi^{a_0}}{\widetilde{g}}
\, D_{a_0}(q^2) \, \gamma_{a_0\pi\eta} \ .
\label{Y a a0}  
\end{eqnarray}
Here $C_\phi^{a_0}$ is given in terms of the coefficients of
Eq.~(\ref{SVV}) and a scalar mixing angle in Eq.~(8) of 
Ref.~\cite{BHS}
and will be considered, for generality, a single parameter.
Furthermore we use the simple $a_0$ propagator:
\begin{equation}
D_{a_0}(q^2) = \frac{1}{ m_{a_0}^2 - q^2 - i m_{a_0} \Gamma_{a_0} }
\ .
\label{aprop}
\end{equation}
Also, $q$ is the positive quantity:
\begin{equation}
q = \left[(p_0 -k_0)^2 -({\bf p}-{\bf k})^2\right]^{1/2}.
\label{pietainvmass}
\end{equation}

Finally, the  SPP type coupling constant in Eq.~(\ref{Y a a0}) as well
as others needed in this paper are defined from the Lagrangian
density:
\begin{eqnarray}
{\cal L}_{SPP} 
&=& -\gamma_{a_o\pi\eta}a_0^0\partial_\mu\pi^0\partial_\mu\eta
-\frac{\gamma_{f_0\pi\pi}}{\sqrt 2}
  f_0\partial_\mu\pi^0\partial_\mu\pi^0
\nonumber\\
&& {}
-\frac{\gamma_{aK{\bar K}}}{\sqrt 2}
  a_0^0\partial_\mu K^-\partial_\mu K^+
-\frac{\gamma_{fK{\bar K}}}{\sqrt 2}
  f_0\partial_\mu K^-\partial_\mu K^+
+\cdots
\ .
\label{SPPLag}
\end{eqnarray}
The relations between these coefficients to  $A$, $B$, $C$, $D$ 
in Eq.~(\ref{LagL0}) are given in Appendix C of
Ref.~\cite{Black-Fariborz-Sannino-Schechter:99}.
The ``$q$-distribution''
$d \Gamma(\phi\rightarrow \pi^0\eta\gamma)/dq$ is
expressed as
\begin{eqnarray}
\frac{ d \Gamma(\phi\rightarrow \pi^0\eta\gamma)}{dq}
&=&
\frac{\alpha}{768\pi^2}
\left( \frac{M_\phi^2 - q^2}{M_\phi} \right)^3
  \sqrt{
    \frac{ \left[ q^2 - (m_\eta+m_\pi)^2 \right]
           \left[ q^2 - (m_\eta-m_\pi)^2 \right]}{ q^2 }
  }
\nonumber\\
&&
{} \times
\left( q^2 - m_\pi^2 - m_\eta^2 \right)^2
\left\vert Y_{a_0}^{(\pi\eta)} \right\vert^2
\ .
\label{pietadistrib}
\end{eqnarray}
 Discussion of the phase space integral is given, for example, 
in Ref.~\cite{PDG}. 

    Now let us see how well we can fit the experimental data on
 the $\pi^0\eta$
invariant mass distribution in this model. We will use the inputs:
$m_{a_0}=984.7 \, \mbox{MeV}$ (from the PDG table~\cite{PDG});
$\Gamma_{a_0} = 70 \, \mbox{MeV} $ (from~\cite{FS});
 $\gamma_{a_0\pi\eta} = -6.80 \, \mbox{GeV}^{-1} $ 
(from ~\cite{FS,BFS}).

\begin{figure}[htbp]
\begin{center}
\epsfbox{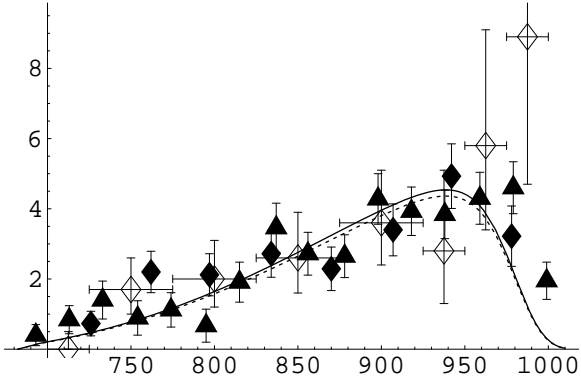}
\end{center}
\caption[]{$d B(\phi\rightarrow \pi^0\eta \gamma) /d q \times
 10^7$ (in units of $\mbox{MeV}^{-1}$) as a function in
 the $\pi^0$-$\eta$ invariant mass $q = m_{\pi^0\eta}$ (in MeV).
 Solid line shows the $a_0$ contribution with
 the best fitted value $C_\phi^{a_0}=3.7\,\mbox{GeV}^{-1}$,
 and the dashed line shows that with
 $C_\phi^{a_0}=3.6\,\mbox{GeV}^{-1}$,
 Experimental data indicated by white diamonds ($\Diamond$)
 are from the SND collaboration in Ref.~\cite{Acha:ep},
 and those by filled triangles 
 and filled diamonds are shown in Ref.~\cite{Achasov-Kisilev}
 extracted from the KLOE collaboration in Ref.~\cite{KLOE:ep}.
}\label{fig:fit a0 1}
\end{figure}

Let us perform two types of fits for obtaining the best value of
$C_\phi^{a_0}$ 
(assuming $\tilde{g}$ to be fixed at the value $4.04$):
\begin{eqnarray}
&&   
\mbox{(I) use the data for all values of $q= m_{\pi^0\eta}$}\ ,
\label{type I}
\\
&&
\mbox{(II) use the data for $m_{\pi^0\eta} \ge 850\,\rm{MeV}$}
\ .
\label{type II}
\end{eqnarray} 
The results are
\begin{eqnarray}
&&
\mbox{(I)} \quad
C_\phi^{a_0} = 3.7 \pm 0.1 \,\mbox{GeV}^{-1}\ ,
\quad
\chi^2/\mbox{d.o.f} = 41/(32-1) \ ,
\nonumber\\   
&&
\mbox{(II)} \quad
C_\phi^{a_0} = 3.6 \pm 0.1 \,\mbox{GeV}^{-1}\ ,
\quad
\chi^2/\mbox{d.o.f} = 32/(17-1) \ .
\label{fit a01}
\end{eqnarray}  
Figure~\ref{fig:fit a0 1} shows the resulting plots of
$d B(\phi\rightarrow \pi^0\eta \gamma) /d q$ together with the
experimental data. Note that, since only the combination 
$\gamma_{a_0\pi\eta}C_\phi^{a_0}/\tilde{g}$ appears
 in our fitting 
procedure, the
best fitted curve will not change even if we allow the values of
$\gamma_{a_0\pi\eta}$ and $\tilde{g}$ to vary.

\begin{figure}[htbp]
\begin{center}
\epsfbox{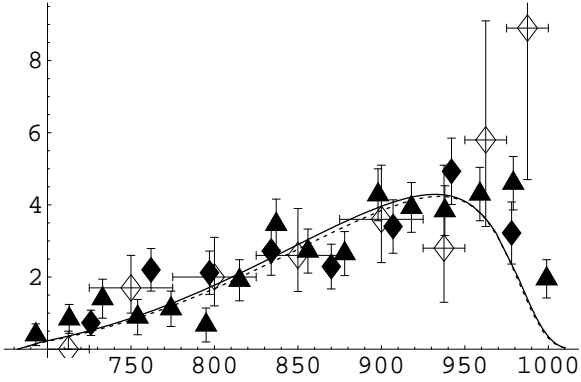}
\end{center}
\caption[]{$d B(\phi\rightarrow \pi^0\eta \gamma) /d q \times
10^7$ (in units of $\mbox{MeV}^{-1}$) as a function in
the $\pi^0$-$\eta$ invariant mass $q = m_{\pi^0\eta}$ (in MeV).
Solid line shows the $a_0$ contribution with
the best fitted values $C_\phi^{a_0}=4.0\,\mbox{GeV}^{-1}$
and $m_{a_0} = 993.2\,\mbox{MeV}$,
and the dashed line shows that with
$C_\phi^{a_0}=3.9\,\mbox{GeV}^{-1}$
and $m_{a_0} = 990.4\,\mbox{MeV}$.
Experimental data are as in Fig.~\ref{fig:fit a0 1}.
}\label{fig:fit a0 2}
\end{figure}

This model %clearly 
gives a poor fit
 to  the experimental
data in the energy region above $950\,\mbox{MeV}$.
One possibility is that the fit may be improved by raising
the mass of $a_0$ above $984.7\,\mbox{MeV}$.
Actually, Ref.~\cite{Acha:ep} gives the best fit value
as $m_{a_0} = 995^{+52}_{-10}\,\mbox{MeV}$.
Let us then fit the $a_0$ mass together with 
value of $C_\phi^{a_0}$.
The results are
\begin{eqnarray}
&&
\mbox{(I)} \quad
C_\phi^{a_0}= 4.0 \pm 0.1 \,\mbox{GeV}^{-1}\ , \quad
m_{a_0} = 993.2 \pm 2.8 \,\mbox{MeV} \ , \quad
\chi^2/\mbox{d.o.f} = 39/(32-2) \ ,
\nonumber\\ 
&&
\mbox{(II)} \quad
C_\phi^{a_0} = 3.9 \pm 0.1 \,\mbox{GeV}^{-1}\ ,\quad
m_{a_0} = 990.4 \pm 2.5 \,\mbox{MeV} \ , \quad
\chi^2/\mbox{d.o.f} = 31/(17-2) \ .
\end{eqnarray}
Note that the best fit value of $m_{a_0}$ in case (II) is
very close to the values shown in Ref.~\cite{Acha:ep}.
In Fig.~\ref{fig:fit a0 2},
we plot 
$d B(\phi\rightarrow \pi^0\eta \gamma) /d q$ together with the
experimental data.
This figure shows that
it is still difficult to reproduce the experimental
data in the energy region above $950\,\mbox{MeV}$ in the present 
model even if one allows the $a_0$ mass to vary.

\begin{figure}[htbp]
\begin{center}
\epsfbox{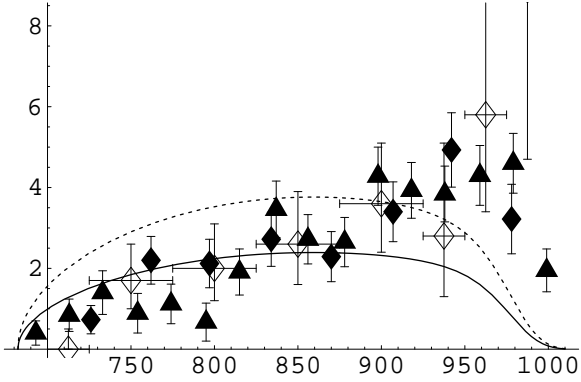}
\end{center}
\caption[]{$d B(\phi\rightarrow \pi^0\eta \gamma) /d q \times
10^7$ (in units of $\mbox{MeV}^{-1}$) as a function in
the $\pi^0$-$\eta$ invariant mass $q = m_{\pi^0\eta}$ (in MeV).
Solid line shows the $a_0$ contribution in the non-derivative
coupling model with
the best fit value $C_\phi^{a_0}=2.13\,\mbox{GeV}^{-1}$
and the dashed line shows that with
$C_\phi^{a_0}=2.68\,\mbox{GeV}^{-1}$
Experimental data are as in Fig.~\ref{fig:fit a0 1}.
}\label{fig:fit a0 l 1}
\end{figure}

For comparison with the chiral symmetric case, we will now
 investigate the effect of 
using non-derivative
coupling at the $a_0\pi^0\eta$ interaction vertex.
This amounts to multiplying Eq.~(\ref{pietadistrib}) by the
factor:
\begin{equation}
\frac{(m_{a_0}^2-m^2_\pi-m^2_\eta)^2}{(q^2-m^2_\pi-m^2_\eta)^2},
\label{nonderivfactor}
\end{equation}
which has the effect of deemphasizing the high $q$ region. It
yields
\begin{eqnarray}
&&
\mbox{(I)} \quad
C_\phi^{a_0} = 2.13 \pm 0.07 \,\mbox{GeV}^{-1}\ , \quad
\chi^2/\mbox{d.o.f} = 113/(32-1) \ ,
\nonumber\\ 
&&
\mbox{(II)} \quad
C_\phi^{a_0} = 2.68 \pm 0.08 \,\mbox{GeV}^{-1}\ ,\quad
\chi^2/\mbox{d.o.f} = 67.9/(17-1) \ .
\label{fit a0l}
\end{eqnarray}
Furthermore, Fig.~\ref{fig:fit a0 l 1}
shows the plot of
$d B(\phi\rightarrow \pi^0\eta \gamma) /d q$ together with the
experimental data.
Comparing this figure with Fig.~\ref{fig:fit a0 1} and
the results in Eq.~(\ref{fit a0l}) with those in
Eq.~(\ref{fit a01}) indicates that the
derivative coupling model gives a better fit.
The non derivative coupling factor clearly seems to 
wash out the resonance peak.

\subsection{$f_0(980)$ production}
\label{ssec:1 f0}

The treatment of the decay $\phi\rightarrow \pi^0\pi^0\gamma$
assuming only the VVS type interaction where S is identified
as $f_0$(980) and subsequently decays to the two neutral
pions, proceeds in a similar manner. Again it is found that
the use of a chiral symmetric derivative type interaction
 is to be preferred
because it does not wash out the scalar resonance peak. However
the overall fit to the $\pi\pi$ invariant mass distribution
is not good, again suggesting that the VVS
type of contribution is not the dominant one.
In this case,   
$d \Gamma(\phi\rightarrow \pi^0\pi^0\gamma)/dq$ is
given by,
\begin{eqnarray}
\frac{ d \Gamma(\phi\rightarrow \pi^0\pi^0\gamma)}{dq}
=
\frac{\alpha}{1536\pi^2}
\left( \frac{M_\phi^2 - q^2}{M_\phi} \right)^3
\sqrt{ q^2 - 4 m_\pi^2 }
\left( q^2 - 2 m_\pi^2 \right)^2
\left\vert Y_{f_0}^{(\pi\pi)} \right\vert^2
\label{phipipispectrum}
\ ,
\end{eqnarray}
where
\begin{eqnarray}
&&
Y_{f_0}^{(\pi\pi)} =
\frac{C_\phi^{f_0}}{\widetilde{g}}
\, D_{f_0}(q^2) \, \sqrt{2} \gamma_{f_0\pi\pi} \ .
\label{Y f f0}  
\end{eqnarray}
The $f_0$ propagator is:  
\begin{equation}
D_{f_0}(q^2) = \frac{1}{ m_{f_0}^2 - q^2 - i m_{f_0} \Gamma_{f_0} }
\ ,
\label{fprop}
\end{equation}
and we will use the mass of the $f_0(980)$ to be ~\cite{HSS}
987 MeV. 
The coupling constant $\gamma_{f_0\pi\pi}$ is related to the width of $f_0$
as~\cite{HSS}
\begin{equation}
\Gamma_{f_0} = \frac{3}{64\pi}
\frac{ \gamma_{f_0\pi\pi}^2}{m_{f_0}}
\sqrt{ 1 - \frac{ 4 m_\pi^2 }{m_{f_0}^2} }
\left( m_{f_0}^2 - 2 m_\pi^2 \right)^2
\ .
\end{equation}
In Ref.~\cite{HSS} a treatment of $\pi\pi$ scattering suggested 
$\Gamma_{f_0} \approx 64.6$\,MeV and correspondingly 
$|\gamma_{f_0\pi\pi}|\approx
2.25\,\mbox{GeV}^{-1}$. 
Considering both $\pi\pi$ and $\pi K$ scattering,
$\gamma_{f_0\pi\pi}\approx 1.47\,\mbox{GeV}^{-1}$ and 
$\Gamma_{f_0}\approx 27.6$\,MeV were determined 
in Ref.~\cite{Black-Fariborz-Sannino-Schechter:99}.

\begin{figure}[htbp]
\begin{center}
\epsfxsize = 6cm
\epsfbox{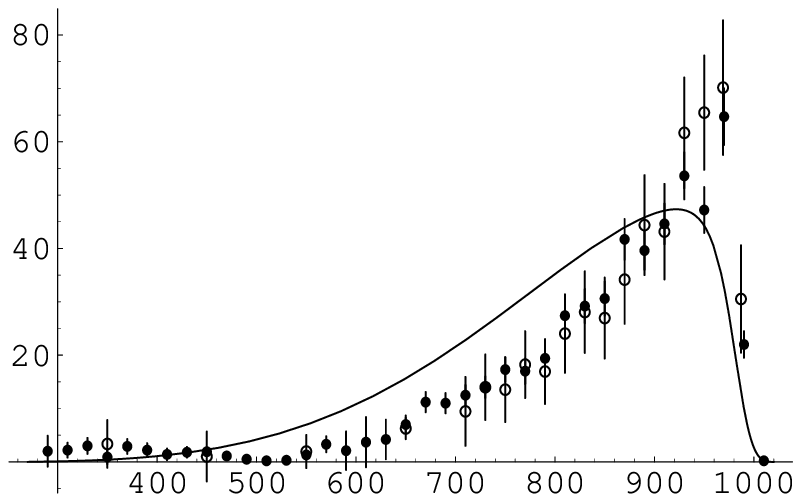}
\
\epsfxsize = 6cm
\epsfbox{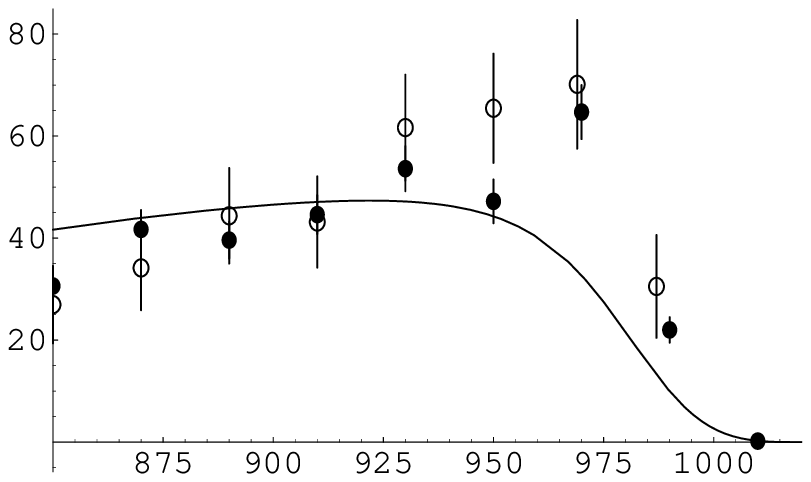}
\\
\hspace*{0cm} \ (a) \ \hspace{4.5cm} \ (b) \ \hspace{3cm}
\end{center}
\caption[]{$d B(\phi\rightarrow \pi^0\pi^0 \gamma) d q \times
10^8$ (in units of $\mbox{MeV}^{-1}$) as a function of
the dipion invariant mass $q = m_{\pi^0\pi^0}$ (in MeV).
Solid line shows the $f_0$ contribution with
$C_\phi^{f_0}=9.3\,\mbox{GeV}^{-1}$.
(a) shows the result in the entire energy region, and
(b) shows that in
$m_{\pi^0\pi^0} \ge 850\,\mbox{MeV}$.
Experimental data shown by $\circ$ are from Ref.~\cite{Acha},
and those by $\bullet$ are from Ref.~\cite{Aloi}.
}\label{fig:fit f0 1}
\end{figure}

Using for example $\vert\gamma_{f_0\pi\pi}\vert = 1.47\,\mbox{GeV}^{-1}$
let us next fit the value of $C_\phi^{f_0}$ to the experimental data.
Furthermore, to avoid any possible confusion with an expected
low energy contribution from the $\sigma$ we shall
use experimental data only in the region
\begin{equation}
m_{\pi^0\pi^0} \ge 850\,\mbox{MeV} \ .
\end{equation}
This yields
\begin{equation}
C_\phi^{f_0} \approx 9.3  \mbox{GeV}^{-1} \ ,
\quad
\chi^2/\mbox{d.o.f.} = 101/(17-1) \ .
\end{equation}
In Fig.~\ref{fig:fit f0 1},
we show the resultant $f_0$ contribution
 together with the
experimental data~\cite{Acha,Aloi}.

\section{Charged $K$ loop contribution}
\label{sec:K-loop}

\begin{figure}[htbp]
\begin{center}
\epsfxsize = 13cm
\epsfbox{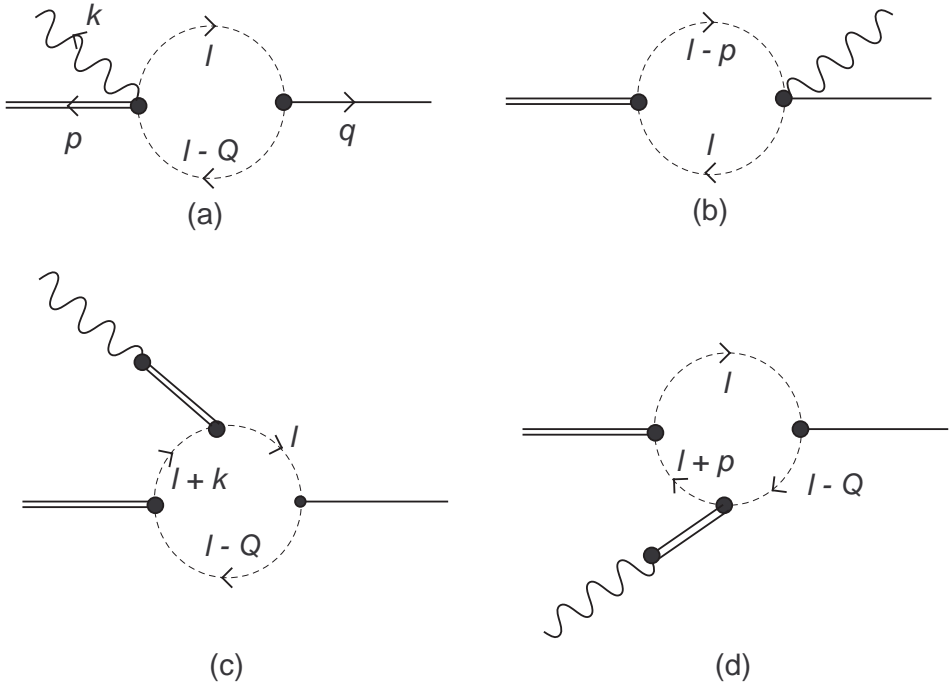}
\end{center}
\caption[]{Feynman diagrams for
the charged $K$ loop contributions to
$\phi(p,\epsilon^V) \rightarrow a_0(Q)+ \gamma(k,\epsilon)$.
The solid line denotes the $a_0$ meson, the wavy line the photon,
the double solid line the vector mesons ($\rho$, $\omega$, $\phi$)
and the dashed line the $K$ meson.
$p$, $k$ and $Q$ are the momenta of the $\phi$ meson, the photon
and the $a_0(980)$, respectively, and
$l$ is the loop momentum.
}\label{fig:Kloop}
\end{figure}

Now let us explore the $K$-loop contributions to the radiative 
$\phi$ decays. The relevant Feynman diagrams are shown in
Fig.~\ref{fig:Kloop}.
The diagrams (c) and (d) each give the same result while (a) and (b)
are required by gauge invariance. Notice from Eq.~(\ref{VMD})
that the direct photon-two pseudoscalar vertex vanishes in this
model when $k=2$ is adopted,
as we are doing here.~\footnote{%
 In the present analysis, we just use $k=2$ for simplicity in
 calculation so 
 that two kaons couple to gamma only through vector meson intermediate
 lines keeping vector meson dominance.
}
Thus the two pseudoscalars first couple to $\rho, \omega$
and $\phi$ which then transform to a photon as shown in Figs.~(c)
 and (d). The strong vector-two pseudoscalar interaction vertices
may be read from the fourth term of Eq.~(\ref{LagL0}) while
the scalar-two pseudoscalar interaction vertices are derived from
the A, B, C and D terms of this equation (and explicitly given 
in Eq.~(\ref{SPPLag})).

Note again that the Lagrangian density of Eq.~(\ref{LagL0})
treats all of the pseudoscalars, scalars and vectors in a consistent
chiral invariant manner. It can be modified to include gauge
invariant photon interactions by making the replacements:
\begin{eqnarray}
v_\mu \rightarrow {\tilde v}_\mu &=& 
  v_\mu +\frac{1}{2}e{\cal A}_\mu
  \left(\xi Q \xi^{\dagger} +\xi^{\dagger}Q\xi \right), 
\nonumber \\
p_\mu \rightarrow {\tilde p}_\mu &=& 
  p_\mu +\frac{1}{2}e{\cal A}_\mu
  \left(\xi Q \xi^{\dagger} -\xi^{\dagger}Q\xi\right),
\nonumber \\           
\rho_\mu \rightarrow {\tilde \rho}_\mu &=& 
  \rho_\mu \ , 
\label{electrify}
\end{eqnarray}
where ${\cal A}$ and $Q$ were defined after Eq.~(\ref{VMD}).
Under an infinitesimal electromagnetic gauge transformation with
$\delta {\cal A}_\mu ={\partial}_\mu \lambda(x)$, 
${\tilde p}_\mu$ and ${\tilde v}_\mu-{\tilde g}{\tilde \rho}_\mu$
 in Eq.~(\ref{electrify}) do not contain any 
terms
proportional to ${\partial}_\mu \lambda(x)$. When substituted
into Eq.~(\ref{LagL0}), the above replacements yield, in addition to
Eq.~(\ref{VMD}) the four field photon interaction terms
 in the Lagrangian density:
\begin{equation}
\frac{em_v^2}{{\tilde g}F_\pi^2}{\cal A}_\mu{\phi_\mu}K^{+}K^{-}
+i\frac{e\gamma_{aKK}}{\sqrt{2}}{\cal A}_{\mu}a_0^0(K^+\partial_\mu 
K^- - K^-\partial_\mu K^+) +\cdots \ ,
\label{fourfield}
\end{equation}
where $\phi_\mu$ is the $\phi$-meson field and $a_0^0$ is the
neutral $a_0(980)$ scalar meson field.~\footnote{
  The terms such as 
  $ \epsilon^{abc} \epsilon_{def} N_a^d
    {(\tilde{g}\rho_\mu-v_\mu)}_b^e {(\tilde{g}\rho_\mu-v_\mu))}_c^f$
  can be added into the Lagrangian (\ref{LagL0}).
  Although they do not contribute at tree level 
  to the radiative decays studied in the present analysis,
  they generate the vertex of type $SV\gamma PP$, which gives
  the quantum correction to $SV\gamma$ vertex.
  Since this quantum correction does not depend on the external
  momenta, its contribution is absorbed into the redefiniton 
  of the effective $SV\gamma$ coupling $C_V^S$.
}
Now it is straightforward
to obtain the $K$ loop amplitudes (with the assumption $k=2$) for
$\phi(p,\epsilon^V) \rightarrow a_0(Q)+ \gamma(k,\epsilon)$:
\begin{eqnarray}
S_a &=& h\int \frac{d^4l}{i(2\pi)^4}\frac{[l\cdot(Q-l)][\epsilon\cdot
\epsilon^V]}{[l^2+m_K^2][(Q-l)^2+m_K^2]},
\nonumber \\
S_b &=&-\frac{h}{2}\int 
\frac{d^4l}{i(2\pi)^4}
\frac{[(2l-p)\cdot\epsilon^V][(2l-p)\cdot\epsilon]}
{[l^2+m_K^2][(p-l)^2+m_K^2]},
\nonumber \\ 
S_c=S_d&=&-\frac{h}{2}\int 
\frac{d^4l}{i(2\pi)^4}\frac{[l\cdot(Q-l)][(2l+k)\cdot\epsilon]
[(2l+k-Q)\cdot\epsilon^V]}{[l^2+m_K^2][(Q-l)^2+m_K^2][(k+l)^2+m_K^2]},
\label{loopamp}
\end{eqnarray}
where
\begin{equation}
h=\frac{em_v^2\gamma_{aKK}}{\sqrt{2}{\tilde g}F_\pi^2} \approx 
{\sqrt{2}}e{\tilde g}\gamma_{aKK},
\label{hcoeff}
\end{equation}
and the KSRF relation was used in the last step.
Note that the quantity defined in Eq.~(\ref{pietainvmass}), 
$q^2=-Q^2$.
 To get the
amplitude for the decay $\phi \rightarrow f_0 \gamma$ we
should replace $\gamma_{aKK}$ by $\gamma_{fKK}$ in 
Eq.~(\ref{hcoeff}).

The next step is to regulate the divergences which occur
in these amplitudes. We employ the dimensional regularization
scheme and
 thus continue from 4 to $d$ space-time dimensions
according to the formula:
\begin{equation}
\int\frac{d^dl}{i(2\pi)^d}\frac{1}{(l^2+s)^n}=\frac{\Gamma(n-d/2)}
{(4\pi)^{d/2}\Gamma(n)s^{n-d/2}},
\label{dimreg}
\end{equation}
where $n$ is an integer while $s$ is arbitrary. 
The physical amplitudes
will emerge in the limit when $\epsilon=4-d \rightarrow 0$. It
is convenient to define:
\begin{equation}
\frac{1}{{\bar \epsilon}}=\frac{2}{\epsilon} -\gamma +
 \ln(4\pi),
\label{epsilonbar}
\end{equation}
where $\gamma \approx$ 0.577
is the Euler-Mascheroni constant.

For $S_a$ we use the identity $l\cdot Q-l^2 =-1/2[(l-Q)^2+m_K^2]
-1/2(l^2+m_K^2) +(m_K^2 + Q^2/2)$ to write:
\begin{equation}
S_a=ih\delta_{\mu\nu}\epsilon^V_\mu(p)\epsilon_\nu(k)[-A_0(m_K^2)
+\frac{1}{2}(2m_K^2 +Q^2)B_0(Q^2)],
\label{Sa}
\end{equation}
where
\begin{eqnarray}
A_0(m_K^2)&=&\int\frac{d^dl}{i(2\pi)^d}\frac{1}{l^2+m_K^2}\ ,
\nonumber \\
B_0(Q^2) &=&\int
\frac{d^dl}{i(2\pi)^d}\frac{1}
{[l^2+m_K^2][(l-Q)^2+m_K^2]} \ . 
\label{A0B0}
\end{eqnarray}

For $S_b$ we define:
\begin{eqnarray}
S_b &=& -i\frac{h}{2}\epsilon^V_\mu(p)\epsilon_\nu(k)B_{\mu\nu}(p),
\nonumber \\
B_{\mu\nu}(p) &=&\int
\frac{d^dl}{i(2\pi)^d}\frac{(2l-p)_\mu(2l-p)_\nu}
{[l^2+m_K^2][(p-l)^2+m_K^2]}\ .
\label{Bmunu}
\end{eqnarray}              

Finally, for the triangle diagrams we similarly rearrange the
numerator to get:
\begin{eqnarray}
S_c=S_d&=&-i\frac{h}{2}\epsilon^V_\mu(p)\epsilon_\nu(k)
[-\frac{1}{2}B_{\mu\nu}(k)
+{Q_\mu}{B_\nu}(-k)+\frac{1}{2}{Q_\mu}{k_\nu}{B_0}(k^2)
\nonumber \\
&-&\frac{1}{2}B_{\mu\nu}(p)-{B_\mu}(p)Q_\nu
+\frac{1}{2}{p_\mu}{Q_\nu}{B_0}(p^2)+\frac{1}{2}(2m_K^2+Q^2)X_{\mu\nu}(p,k)],
\label{triangle}
\end{eqnarray}
wherein $B_0(p^2)$ and $B_{\mu\nu}(p)$ have been
already defined while,
\begin{equation}
B_\mu(p)=\int\frac{d^dl}{i(2\pi)^d}\frac{l_\mu}
{[l^2+m_K^2][(l-p)^2+m_K^2]},
\label{Bmu}
\end{equation}
and
\begin{equation}
X_{\mu\nu}(p,k)=
\int\frac{d^dl}{i(2\pi)^d}\frac{(2l+k-Q)_\mu(2l+k)_\nu}{[l^2+m_K^2]
[(Q-l)^2+m_K^2][(k+l)^2+m_K^2]}.
\label{Xmunu}
\end{equation}
Note that $k^2=0$ since it corresponds to a physical photon momentum.

Using Feynman's trick for combining denominators
 and Eqs.~(\ref{dimreg}) and (\ref{epsilonbar})
 we evaluate the integral $B_0(p^2)$ near $d=4$:
\begin{eqnarray}
B_0(p^2)&=&\frac{1}{(4\pi)^2}[\frac{1}{{\bar \epsilon}}-F_0(p^2)],
\nonumber \\
F_0(p^2)&=&\int^1_0 dx \ln[m_K^2-x(x-1)p^2].
\label{evaluateB0}
\end{eqnarray}
We also find
\begin{equation}
B_\mu(p)=\frac{1}{2}p_\mu B_0(p^2).
\label{evaluateBmu}
\end{equation}
The presence of a pole at $d=4$ indicated by the term $1/{\bar \epsilon}$ 
corresponds, of course, to a logarthmic divergence
in the cutoff regularization method.

Using the integrals defined above we can compactly write
the total amplitude as:
\begin{eqnarray}
S &=& S_a+S_b+S_c+S_d
\nonumber \\
 &=& i\frac{h}{2}\epsilon^V_\mu(p)\epsilon_\nu(k)[\delta_{\mu\nu}
(-2A_0(m_K^2)+(2m_K^2 + Q^2)B_0(q^2)) +B_{\mu\nu}(k)
\nonumber \\ 
&-&(2m_K^2 + Q^2)X_{\mu\nu}
(p,k)].
\label{totalamp}
\end{eqnarray}
Notice, in particular, that the contribution of $S_b$ has cancelled
out against a piece of the triangle 
diagrams.

The evaluation of an integral of the form $B_{\mu\nu}(p)$
is a little more complicated. We use covariance (in $d$-dimensions)
to relate it to the other integrals as:
\begin{eqnarray}
B_{\mu\nu}(p) &=& \delta_{\mu\nu}\frac{4}{1-d}[-\frac{A_0(m_K^2)}{2}
+(m_K^2+\frac{p^2}{4})B_0(p^2)] \nonumber \\
&+&{p_\mu}p_\nu [\frac{4}{p^2}((1+\frac{1}{2}\frac{d}{1-d})A_0(m_K^2)
-m_K^2(1+\frac{d}{1-d})B_0(p^2)
\nonumber \\
&-&\frac{p^2}{4}\frac{d}{1-d}B_0(p^2))-B_0(p^2)].
\label{evaluateBmunu}
\end{eqnarray}
We see that $B_{\mu\nu}$ contains the integral $A_0(m_K^2)$ which is
noted from 
Eq.~(\ref{A0B0}) to involve a quadratic divergence in the cut-off
regularization 
scheme. In the dimensional regularization approach this 
corresponds~ \cite{vandhy}
to a pole at $d=2$, as may be seen from Eq.~(\ref{dimreg}). 
It is necessary to
check that this divergence cancels out in the total amplitude.
This may
be done by using Eq.~(\ref{evaluateBmunu}) to get, near $d=2$:
\begin{equation}
B_{\mu\nu}(k)|_{k^2=0} = 2A_0(m_K^2)\delta_{\mu\nu} +\cdots\ ,
\label{2dBmunu}
\end{equation}
where the three dots indicate terms not containing $A_0(m_K^2)$.
Substituting this into Eq. (\ref{totalamp}) (considered at d=2)
shows that all dependence on $A_0(m_K^2)$ at $d=2$ is cancelled, as 
desired. We interpret this as the cancellation of the   
quadratic divergences in the individual diagrams.

For the physical case we must consider, of course, the amplitude
evaluated near $d=4$. The integral $A_0(m_K^2)$ is, near $d=4$:
\begin{equation}
A_0(m_K^2)= -\frac{m_K^2}{(4\pi)^2}
  \left[\frac{1}{{\bar \epsilon}}+1-\ln(m_K^2)\right]\ .
\label{evaluateA0}
\end{equation}
Using Eq.~(\ref{evaluateBmunu}) we find for $B_{\mu\nu}(k)$ near 
$d=4$ and $k^2=0$:
\begin{equation}
B_{\mu\nu}(k)=2A_0(m_K^2)\delta_{\mu\nu}+
\frac{1}{3(4\pi)^2}k_\mu k_\nu \left[\frac{1}{{\bar 
\epsilon}}-\ln(m_K^2)+\frac{2}{3}\right]\ ,
\label{4dBmunu}
\end{equation}
wherein the first term was separated for convenience. Note that the
${k_\mu}k_\nu$ term will not contribute to the physical amplitude
because it gets multiplied by the photon polarization vector
 $\epsilon_\nu (k)$. Now substituting Eq.~(\ref{4dBmunu}) into
the total amplitude,  Eq.~(\ref{totalamp}), 
shows that its effect is simply
to cancel the $-2A_0(m_K^2)$ term.

To evaluate the remaining, $X_{\mu\nu}(p,k)$ term we first use
covariance to express it as:
\begin{equation}
X_{\mu\nu}(p,k)=\delta_{\mu\nu}X_1 + p_{\mu}p_{\nu}X_2 
+ k_{\mu}k_{\nu}X_3 
+p_{\mu}k_{\nu}X_4 + k_{\mu}p_{\nu}X_5\ ,
\label{X}
\end{equation}
where each of the $X_i$ depends on $p^2$ and $p{\cdot}k$.
The $X_i$ may be determined by calculating $X_{\mu\mu}$, 
$k_{\mu}X_{\mu\nu}$,
$p_{\nu}X_{\mu\nu}$ and $k_{\mu}p_{\nu}X_{\mu\nu}$ both 
from Eq.~(\ref{X})
and from Eq.~(\ref{Xmunu}). This leads to the relations (remembering
$k^2=0$):
\begin{eqnarray}
X_2&=&0, \nonumber \\
B_0(Q^2)&=&X_1+k{\cdot}pX_5, \nonumber \\
0&=&k{\cdot}pX_3 +p^2X_4, \nonumber \\
X_5&=&\frac{1}{(p{\cdot}k)^2}[\frac{p^2}{d-2}(B_0(Q^2)-B_0(p^2))+
(p{\cdot}k + \frac{Q^2-p^2}{d-2})B_0(Q^2)
\nonumber \\
&+& \frac{4m_K^2 k{\cdot}p}{d-2}
C(p^2,k{\cdot}p)],
\label{theXi}
\end{eqnarray}
where the finite integral $C(p^2,k{\cdot}p)$ is given by:
\begin{equation}
C(p^2,k{\cdot}p)=\int\frac{d^4l}{i(2\pi)^4}\frac{1}{[l^2+m_K^2]
[(l-Q)^2+m_K^2][(l+k)^2+m_K^2]}.
\label{C}
\end{equation}
Actually, only the coefficients $X_1$ and $X_5$ remain after $X_{\mu\nu}$
is multiplied by the polarization vectors of the photon and $\phi$ 
meson; furthermore these two  coefficients are related as above. 
Substituting back 
into the total amplitude, Eq.~(\ref{totalamp}) and making use of the 
cancellation between the $A_0(m_K^2)$ and $B_{\mu\nu}(k)$ terms 
discussed before, yields:
\begin{eqnarray}
S&=&
i\,\frac{h}{2}\,\epsilon^V_\mu(p)\epsilon_\nu(k)
\left(-\frac{\delta_{\mu\nu}}{p{\cdot}k}
  +\frac{k_{\mu}p_{\nu}}{(p{\cdot}k)^2}\right)
  (2m_K^2+Q^2)
\Biggl[\frac{p^2}{2}\left\{B_0(Q^2)
-B_0(p^2)\right\}
\nonumber \\
&&{}
+2m_K^2(p{\cdot}k)C(p^2,p{\cdot}k) -\frac{p{\cdot}k}{(4\pi)^2}
\Biggr]\ .
\label{finiteamp}
\end{eqnarray}
Note that the last term arises from the $1/\epsilon$ term in
$B_0(Q^2)$
multiplying the leading $\epsilon$ term of its factor.
{}From Eq.~(\ref{evaluateB0})  we see that the logarithmic divergences
cancel out of the difference $(B_0(Q^2)-B_0(p^2))$. Thus the final 
amplitude is completely finite; both the logarithmic and quadratic
divergences
have been seen to cancel using regularized expressions for
everything. The quadratic divergences arose in the first place
because of the derivative-type interactions required by use of the
non linear sigma model terms to describe the pseudoscalar meson 
interactions. In addition, the starting Lagrangian treated the vector
and scalar mesons in the same chiral invariant framework.

Evaluation of the finite integrals in Eq.~(\ref{finiteamp}) yields the
final 
expression for the Feynman amplitude, $iS$:
\begin{eqnarray}
iS&=&-\frac{h}{2}\frac{1}{(4\pi)^2}\epsilon^V_\mu(p)\epsilon_\nu(k)
(-\frac{\delta_{\mu\nu}}
{p\cdot k}+\frac{k_{\mu}p_{\nu}}{(p{\cdot}k)^2})(2m_K^2+Q^2)
%\left{\{}
\biggl\{
-p{\cdot}k
\nonumber \\
&+&\frac{p^2}{2}\left(\sigma(p^2)[\ln\frac{1+\sigma(p^2)}
{1-\sigma(p^2)}
-i\pi]-\sigma(Q^2)[\ln\frac{1+\sigma(Q^2)}{1-\sigma(Q^2)}-i\pi]\right)
\nonumber \\
&+&\frac{m_K^2}{2}\left([\ln\frac{1+\sigma(p^2)}{1-\sigma(p^2)}
-i\pi]^2-[\ln\frac{1+\sigma(Q^2)}{1-\sigma(Q^2)}-i\pi]^2\right)
%\right{\}},
\biggr\},
\label{finalamp}
\end{eqnarray}
where,
\begin{eqnarray}
\sigma(p^2)&=&\sqrt{1+\frac{4m_K^2}{p^2}},
\nonumber \\
\sigma(Q^2)&=&\sqrt{1+\frac{4m_K^2}{Q^2}},
\label{sigmas}
\end{eqnarray}
Note that Eq.~(\ref{finalamp}) holds only in the kinematical range
where 
$-Q^2=q^2>4m_K^2$; the positive quantity, $q$ was also defined
 in Eq.~(\ref{pietainvmass}).
Furthermore note that $p^2=-m_{\phi}^2$. In the kinematical range where
$-Q^2=q^2<4m_K^2$, one should replace 
\begin{eqnarray}
[\ln\frac{1+\sigma(Q^2)}{1-\sigma(Q^2)}-i\pi] 
&\rightarrow&-2i\tan^{-1}\frac{1}{{\tilde \sigma}
(Q^2)},
\nonumber \\
\sigma(Q^2)\rightarrow i{\tilde \sigma}(Q^2) 
&=&i\sqrt{-1-\frac{4m_K^2}{Q^2}}\ , 
\label{replacement}
\end{eqnarray}
in Eq.~(\ref{finalamp}) above.

\section{Comparing the $K$ loop with experiment}
\label{sec:comp}

The expression in Eq.~(\ref{finalamp}) describes
 the decay $\phi\rightarrow a_0\gamma$.
To get the Feynman amplitude for $\phi\rightarrow \pi^0\eta\gamma$,
 we should multiply Eq.~(\ref{finalamp}) by the factor
$(q_1\cdot q_2)D_{a_0}(q^2)\gamma_{a_0\pi\eta}$, where 
$D_{a_0}(q^2)$ was defined in Eq.~(\ref{aprop}). This assumes
a simple form for the $a_0$
propagator, which can only be an approximation. However
our main concern here is an initial exploration 
of the resonance region in the present framework so
it seems reasonable for now. We write the resulting
Feynman amplitude as
\begin{equation}
iS(\phi \rightarrow \pi^0\eta\gamma)=
e(q_1\cdot q_2)X^{(\pi\eta)}_{a_0}
\left[(p\cdot k)(\epsilon^V\cdot\epsilon)-
(p\cdot\epsilon)(k\cdot\epsilon^V)\right]\ ,
\label{Xfpp}
\end{equation}
which thereby defines $X^{(\pi\eta)}_{a_0}$. Note that the
sum $X^{(\pi\eta)}_{a_0}+Y^{(\pi\eta)}_{a_0}$, 
where $Y^{(\pi\eta)}_{a_0}$ is defined in Eq.~(\ref{Y a a0}),
 would correspond
to a model containing both the $K$ loop contribution to the resonant
 amplitude as well
as a point vertex contribution to the resonant amplitude.
For now we focus on the $K$-loop contribution. The decay spectrum
shape, 
$d\Gamma/dq$ is then obtained by replacing $Y^{(\pi\eta)}_{a_0}$ in
 Eq.~(\ref{pietadistrib}) by $X^{(\pi\eta)}_{a_0}$. Conventionally,
one uses instead,
\begin{equation}
\frac{ dB(\phi\rightarrow \pi^0\eta\gamma)}{dq}=
\frac{1}{\Gamma(\phi)}
\frac{ d \Gamma(\phi\rightarrow \pi^0\eta\gamma)}{dq}\ ,
%\label{}
\end{equation}
where $\Gamma(\phi)= 4.26$\,MeV.

\begin{figure}[htbp]
\begin{center}  
%\epsfbox{morekloop1.eps}
%\rotatebox{270}{
%\includegraphics[height=3in,width=3in]{morekloop1.eps}
%}
%\\
\epsfbox{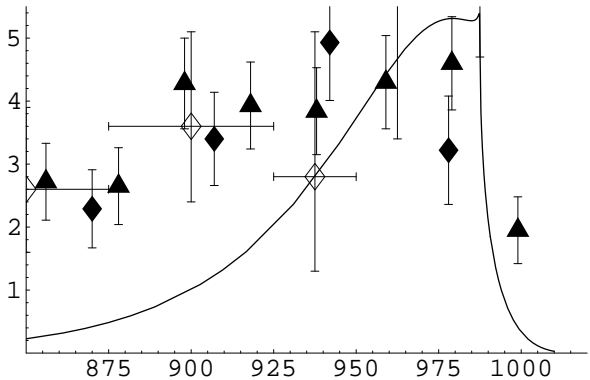}
\end{center}
\caption[]{Predicted $dB(\phi\rightarrow \pi^0\eta\gamma)/dq$ in the
 region of the $a_0(980)$ resonance with
 $\gamma_{aKK}\gamma_{a\pi\eta}= 125\,\mbox{GeV}^{-2}$ 
 and $\Gamma_a = 0.1 $\,GeV. 
 The vertical scale has units $10^{-7}\,\mbox{MeV}^{-1}$.
% while the uncertainties 
% in the experimental points shown (boxes) are about 
% $\pm 0.6 \times 10^{-7} \,\mbox{MeV}^{-1}$.
 Experimental data are as in Fig.~\ref{fig:fit a0 1}.
%\par
% {\bf(M.H.) I made a plot using the same parameters which Joe used.
% However, the experimental data seem different.
% I would appreciate if you could let me know which data are used.
%(J.S.) The data I used were the KLOE and SND data
% quoted by Achasov in ref 24. Also, I
%think that Masa's plots look nicer than mine do.  }
}
\label{fig:morekloop1}
\end{figure}

    In section~\ref{ssec:1 a0} we observed that, even though
 the use of derivative type SPP coupling helped somewhat, the tree 
interaction involving the $a_0(980)$ resonance 
was unable to explain the shape of the
peak at large $q$ in the experimental data for
$dB(\phi\rightarrow \pi^0\eta\gamma)/dq$. Now we will look at the
result of using the $K$-loop amplitude for this purpose.
Taking~\cite{PDG} $m_a = 985$\,MeV and $\Gamma_a =50$-$100$\,MeV,
the 
only quantity which is not well known is the product of the 
scalar meson coupling constants $\gamma_{aKK}\gamma_{a\pi\eta}$.
In Fig.~\ref{fig:morekloop1}, it is shown that a choice
$\gamma_{aKK}\gamma_{a\pi\eta}= 125 \,\mbox{GeV}^{-2}$ 
can nicely explain
the shape of the experimental data in the region of $q$ near the 
$a_0(980)$ resonance. 

For $q$ below the resonance region, the $K$ loop
contribution in the present model falls off rapidly, as one might
reasonably expect with derivative coupling, and lies lower
 than the data points.
In addition to the nonresonant background~\cite{Achasov-Ivanchenko}  
which is usually included to explain this region, there might be some
tree level
resonance production which was observed in Fig.~\ref{fig:fit a0 1}
to peak around $950$\,MeV.

\begin{figure}[htbp]
\begin{center}
%\epsfbox{morekloop1a.eps}
%\includegraphics[height=3in,width=3in]{morekloop1a.eps}
%\rotatebox{270}{
%\includegraphics[height=3in,width=3in]{morekloop1a.eps}
%}
%\\
\epsfbox{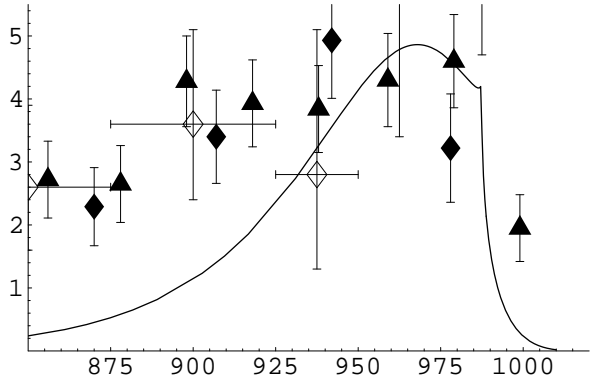}
\end{center}
\caption[]{Predicted $dB(\phi\rightarrow \pi^0\eta\gamma)/dq$
 in the region of the $a_0(980)$ resonance but where
 the $a_0$ mass was artificially lowered to $970$\,MeV.
 Here  $\gamma_{aKK}\gamma_{a\pi\eta}= 115\,\mbox{GeV}^{-2}$
 and $\Gamma_a = 0.1$\,GeV.
 Experimental data are as in Fig.~\ref{fig:fit a0 1}.
%\par
% {\bf(M.H.) I made a plot using the same parameters which Joe used.
%(J.S.) It also looks nicer than mine.}
}
\label{fig:morekloop1a}
\end{figure}

   The main feature of the data is that there is a very
 rapid falloff with apparent discontinuity
of the slope, when $q$ reaches the $K {\bar K}$ threshold. This
is a clear signal for the importance of the $K$ loop
contribution. One may see this feature
 by referring to Fig.~\ref{fig:morekloop1a},
for which the $a_0(980)$ mass has been artificially lowered to 
$970$\,MeV. Comparing with the previous figure shows that the
sharp fall-off is exactly the same in both cases, clearly
unaffected by the difference of assumed resonance masses in 
the two cases. The difference in masses,
on the other hand, shows up as a difference in position
of the peaks. It should be remarked that the peak position
is also affected by the decreasing phase space with
increasing $q$. This characteristic feature of the $K$-loop
contribution was first illustrated by Achasov~\cite{A03} 
by considering the behavior of the result with lowered values
of the $K$-meson mass.

\begin{figure}[htbp]
\begin{center}
\epsfbox{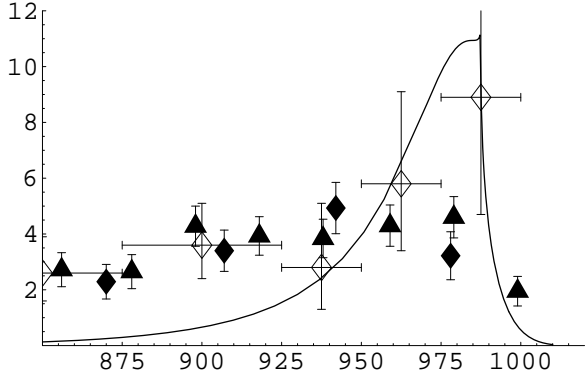}
\end{center}
\caption[]{Predicted $dB(\phi\rightarrow \pi^0\eta\gamma)/dq$ in the
 region of the $a_0(980)$ resonance with
 $\gamma_{aKK}\gamma_{a\pi\eta}= 95\,\mbox{GeV}^{-2}$ 
 and $\Gamma_a = 0.05$\,GeV.
 Experimental data are as in Fig.~\ref{fig:fit a0 1}.
}
\label{fig:morekloop1b}
\end{figure}

%\footnote{
% \bf(M.H.) I studied the dependence of the theoretical curve
% on the width of $a_0(980)$.
% What do you think ? (J.S.) It seems fine; I changed the English a
%little. 
%}
Next, let us check the dependence of the prediction on the
width of $a_0(980)$.
Figure~\ref{fig:morekloop1b} shows that the predicted
$dB(\phi\rightarrow \pi^0\eta\gamma)/dq$ in the
region of the $a_0$(980) resonance with
$\gamma_{aKK}\gamma_{a\pi\eta}= 95 \mbox{GeV}^{-2}$ 
and $\Gamma_a = 0.05$\,GeV.
Comparing this prediction with that given in 
Fig.~\ref{fig:morekloop1}, we see that the smaller $a_0$ width
gives a sharper peak, and that a smaller value of 
$\gamma_{aKK}\gamma_{a\pi\eta}$ can also reproduce
the experimental data at the peak position.
For further decreasing the value of $\gamma_{aKK}\gamma_{a\pi\eta}$
the
%~\footnote{
% \bf(M.H.)
% I am not sure that this statement is correct.
% What do you think ? (J.S.) I changed the statement a
%little bit because I think the mechanism of including
%the K loop in the $a_0$ propagator is a little different.
%It seems to be like the Flatte mechanism. When one looks
%below the k k bar threshold there is a kinematical
% real part which
%changes the pole location. Above the threshold there is
% a kinematical imaginary
%part which changes the width. The net result seems to be a 
%cusp shape rather than a Breit Wigner shape. It is also
%effctively narrower. Before going into detail
%on this I would like to understand better the two channel
%pi pi, K Kbar scattering and the analogous mechanism
%for the $f_0(980)$ using the K-matrix type unitarization.
%But I think that is better left for a future paper. 
%}
 inclusion of the $K$-loop correction
into the propagator of $a_0(980)$ may be important 
as pointed out in Ref.~\cite{A03}.

    The $K$-loop contribution to the branching distribution,
$dB(\phi\rightarrow \pi^0\pi^0\gamma)/dq$ may be similarly
evaluated and compared to experiment. There is similarly a problem
for the tree level resonance model to reproduce this experimental
shape in the high $q$ region. The $K$ loop amplitude 
$\phi\rightarrow f_0(980)\gamma$ is given by 
Eq.~(\ref{finalamp}) wherein the overall
factor $h$ is now obtained by replacing $\gamma_{aKK}$ in 
Eq.~(\ref{hcoeff}) by $\gamma_{fKK}$. 
To get the Feynman amplitude for 
$\phi\rightarrow \pi^0\pi^0\gamma$,
 we should multiply Eq.~(\ref{finalamp}) by the factor
$\sqrt{2}(q_1\cdot q_2)D_{f_0}(q^2)\gamma_{f_0\pi\pi}$, where
$D_{f_0}(q^2)$ was defined in Eq.~(\ref{fprop}). This
defines $X^{(\pi\pi)}_{f_0}$ as in Eq.~(\ref{Xfpp}).
The spectrum shape is determined by using 
Eq.~(\ref{phipipispectrum}) with 
$X^{(\pi\pi)}_{f_0}$ replacing  $Y^{(\pi\pi)}_{f_0}$ .

\begin{figure}[htbp]
\begin{center}
%\epsfbox{morekloop2.eps}
%\includegraphics[rotate=90,height=3in,width=3in]{morekloop2.eps}
%\rotatebox{270}{
%\includegraphics[height=3in,width=3in]{morekloop2.eps}
%}
\epsfbox{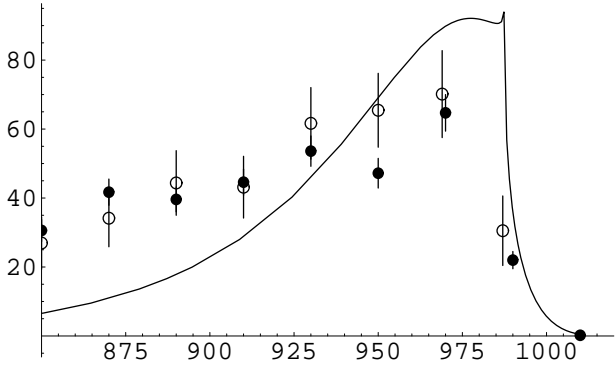}
\end{center}
\caption[]{Predicted $dB(\phi\rightarrow \pi^0\pi^0\gamma)/dq$
 in the region of the $f_0(980)$ resonance
 with $\gamma_{fKK}\gamma_{f\pi\pi}= 86\,\mbox{GeV}^{-2}$
 and $\Gamma_f = 0.1$\,GeV. The vertical scale
 has units $ 10^{-8} \mbox{MeV}^{-1}$.
% while the uncertainties in the
%experimental points shown (boxes) are about $\pm$ 1.0
%$\times 10^{-7} MeV^{-1}$.
}
\label{fig:morekloop2}
\end{figure}

Taking~\cite{PDG} 
$m_f = 980 \pm 10$\,MeV and $\Gamma_f =40$-$100$\,MeV, the
only quantity which is not well known is the product of the
scalar meson coupling constants $\gamma_{fKK}\gamma_{f\pi\pi}$.
In Fig.~\ref{fig:morekloop2}, it is shown that a choice
$\gamma_{aKK}\gamma_{f\pi\pi}= 86 \,\mbox{GeV}^{-2}$ 
can nicely explain
the shape of the experimental data in the region of $q$ near the
$f_0(980)$ resonance.                                                 

    As in the case of the $\phi\rightarrow\pi^0\eta\gamma$
process, the  $K$-loop description of $dB/dq$ only explains 
the upper $q$ region near the scalar resonance. To cover the 
lower $q$
region some non resonant background~\cite{Achasov-Ivanchenko}
is required. Possibly a tree level resonant background,
corresponding to using
 $X^{(\pi\pi)}_{f_0} +Y^{(\pi\pi)}_{f_0}$ in 
Eq.~(\ref{phipipispectrum}), would also be appropriate.

     In both cases considered in this section, it is also
 desirable to include the effects of using multichannel
scalar meson propagators \cite{multichaprop} for a better
approximation to the detailed dynamics. A very recent treatment
of the $\phi\rightarrow\pi^0\pi^0\gamma$ process in this
framework is given in \cite{AK2005}.

\section{Summary and discussion}
\label{sec:SD}

    Historically, the study of elementary particle spectroscopy
has been built around the organization of these particles into
SU(3) flavor multiplets and the consequent predicted (broken symmetry)
mass formulas and interaction vertices. The still mysterious
scalars can be expected to yield up some of their secrets by this
type of analysis. Indeed some recent analyses have already been
carried out~\cite{Black-Fariborz-Sannino-Schechter:99,%
cenp,oller,pelaez,mppr}.
The most dramatic feature is that the light scalars appear to exhibit,
as originally suggested by Jaffe~\cite{j},
a reverse mass ordering compared to the other meson multiplets. 

    In Ref.~\cite{BHS} an attempt was made to extend the SU(3)
analysis 
to relate all the decays of the types S$\rightarrow
\gamma\gamma$, V$\rightarrow$S$\gamma$ and  S$\rightarrow$V$\gamma$
to each other by using a simple effective VVS point-like interaction,
 together with vector meson dominance.
The analogous assumption~\cite{gsw} of a
point-like VVP structure was
very successful~\cite{pod} in phenomenologically correlating  
P$\rightarrow\gamma\gamma$, 
V$\rightarrow$P$\gamma$ and   P$\rightarrow$V$\gamma$
decays. Such an approach was the original motivation which led
 to the present investigation. In section 2
we compared the spectrum shape of the decays 
$\phi\rightarrow\eta\pi^0$,
measuring the effects of an intermediate  $a_0(980)$ 
resonance and $\phi\rightarrow \pi^0\pi^0$, 
measuring the effects of an
intermediate $f_0(980)$ resonance, in the point like VVS model
with the corresponding experimental
observations. It was found that the resonant peaks in the model were  
pushed lower due to  decreasing phase space. This contrasted with 
experiment which does not indicate this effect. On the other hand,
if one were to use a tree model of this type with non derivative
SPP type couplings, the resonant peaks were seen
to get completely washed out. This would appear to be an advantage
for the derivative coupling, which is dictated by chiral symmetry in
the  
present framework. Nevertheless, since even with derivative coupling
the spectrum shape is not very well fitted, there must be another 
mechanism at work.

    Now, it has been emphasized~\cite{A03} that the $K$-loop model
for the $\phi$ radiative decays constitutes a special mechanism
which
 does give a characteristic spectrum shape in agreement  
with experiment. This is readily understandable since the 
$\phi(1020)$
meson is just a little bit heavier than the two $K$ mesons which
comprise its main decay product.
 We thus studied the $K$-loop diagrams using
the chiral Lagrangian of pseudoscalars, vectors and scalars
given in Eqs.~(\ref{LagL0}) and (\ref{SVV}) with the relevant photon
terms introduced by the substitutions shown in Eq.~(\ref{electrify}).
Most of the calculations of this process have not started from a 
chiral symmetric Lagrangian and have thus used non-derivative type
SPP type interaction vertices. The use of derivative coupling
introduces an extra complication in that there is an new
diagram, shown as (b) in Fig.~\ref{fig:Kloop}. 
In addition, individual
diagrams now contain quadratic as well as logarithmic divergences.
It is known that these divergences are forced to cancel from
gauge invariance. However we have used the dimensional regularization
scheme and shown explicitly in 
section~\ref{sec:K-loop} that both the log and
quadratic 
divergences cancel in the regularized expressions. This
may be of some interest in dealing with processes
of the present type.   

In section~\ref{sec:comp}, 
we observed that the shape of the $a_0(980)$
and $f_0(980)$ resonance regions in the $\phi$ radiative decays
could be explained by the corresponding $K$-loop amplitudes.
Furthermore, it was evident that the characteristic sharp
drop in the amplitude at large $q^2$ was associated with the 
the $K {\bar K}$ threshold rather than with the falloff of the
resonance away from its peak. For this work, we used the
coupling constant products $\gamma_{aKK}\gamma_{a\pi\eta}$   
and $\gamma_{fKK}\gamma_{f\pi\pi}$ respectively
as fitting parameters for the $\phi$ radiative decay spectra into
$\pi^0\eta$ and $\pi^0\pi^0$. Elsewhere, we plan to study 
more precisely the values of these coupling constants
obtained by comparing with experiment, chiral models of
meson meson scattering in which the same 
interactions are used. We also will study how the point-like
diagrams with resonant contributions discussed 
in section~\ref{sec:VVS} can be used in 
conjunction with the $K$-loop diagrams to improve the fit
to the resonant region. 
 This will presumably become even more
interesting when more data points become available.
Another point of interest concerns the extent to which the
various SPP coupling constants can be correlated assuming a single
nonet of scalars. This arises because there is some 
evidence~\cite{mixing} that
two scalar nonets (one presumably made from 4 quarks and the other
from two quarks) mix to make up the physical scalar states. 
A recent exploration of the effect of such a mixture on $\phi(1020)$
radiative decays has been given in Ref.~\cite{tkm}.
Still another correction to the simple picture employed here would
be to use more realistic resonance propagators by including  
pseudoscalar loops~\cite{multichaprop}.

    Of course, in order to make a careful comparison with experiment
one should include non-resonant contributions which are expected
to dominate for small $q$. These will include 
the emission of a pion with a virtual $\rho$ which subsequently
decays into  $\pi\gamma$ 
(and similar diagrams leading to a $\pi^0\eta \gamma$ final state)
as discussed in Ref.~\cite{Achasov-Ivanchenko}.  
There will also be non resonant contributions from the $K$-loop
diagrams. A variety of interference mechanisms to explain the 
full spectrum are discussed in Ref.~\cite{gksy}. 
It should be noted that
the ``background" contributions may very well have a non-trivial
effect also in the resonance region itself.

Using the results obtained here and taking into account the features
just discussed, we will continue to 
study the $\phi$ radiative decays with the expectation that it
may contribute to the understanding of the puzzling  scalar mesons
and ultimately to low energy QCD. 

DEDICATION: We are pleased to dedicate this paper to Rafael
Sorkin in connection with the world wide web celebration of
his sixtieth birthday. We have benefited a great deal from our
interactions with Rafael. We wish him good health and continued
success in his endeavor to understand the deepest mysteries of
space-time structure.

\section*{Acknowledgments}
We would like to thank 
A.~Abdel-Rehim, N.~N.~Achasov, A.~H.~Fariborz, and 
F.~Sannino for very helpful discussions.
%The work of D.~B. ...
The work of M.~H. is 
supported in part by the Daiko Foundation \#9099, 
the 21st Century
COE Program of Nagoya University provided by Japan Society for the
Promotion of Science (15COEG01), and the JSPS Grant-in-Aid for
Scientific Research (c) (2) 16540241.
The work of J.~S. is supported in part by the U. S. DOE
under contract No. DE-FG-02-85ER 40231.

\end{document}